\documentclass[12pt,a4paper]{article} 
\usepackage{amssymb}
\usepackage{graphicx}

\def\BI{\begin{itemize}}
\def\EI{\end{itemize}}
\def\BC{\begin{center}}
\def\EC{\end{center}}
\def\BE{\begin{equation}}
\def\EE{\end{equation}}
\def\BA{\begin{eqnarray}}
\def\EA{\end{eqnarray}}
\def\BAN{\begin{eqnarray*}}
\def\EAN{\end{eqnarray*}}

\title{Turbulent patterns in wall-bounded flows:\\
a Turing instability?}

\author{\Large Paul Manneville\\
\large LadHyX, Hydrodynamics Laboratory, CNRS UMR7646\\
\'Ecole Polytechnique 91128 Palaiseau, France}
\date{\normalsize to appear in Europhysics Letters} 

\begin{document}

\maketitle

\begin{abstract}
In their way to/from turbulence, plane wall-bounded flows display an interesting transitional regime where laminar and turbulent oblique bands alternate, the origin of which is still mysterious.
In line with Barkley's recent work about the pipe flow transition involving reaction-diffusion concepts, we consider plane Couette flow in the same perspective and transform Waleffe's classical four-variable model of self-sustaining process into a reaction-diffusion model.
We show that, upon fulfillment of a condition on the relative diffusivities of its variables, the featureless turbulent regime becomes unstable against patterning as the result of a Turing instability. A reduced two-variable model helps us to delineate the appropriate region of parameter space. An {\it intrinsic\/} status is therefore given to the pattern's wavelength for the first time. Virtues and limitations of the model are discussed, calling for a microscopic support of the phenomenological approach.\\[1ex]
\end{abstract}

\section{The context}
\label{s1}

Patterns are currently observed in continuous media driven out of equilibrium \cite{CH93}, natural convection being an emblematic case.
Pattern formation is indeed often an obliged stage in the transition to turbulence.
In this respect, wall-bounded shear flows are systems of great theoretical and practical interest.
Contrasting with  bulk shear flows that become turbulent at low Reynolds numbers%
\footnote{Comparing shear effects to viscous dissipation over some relevant scale, the Reynolds number ($R$ in the following) is the natural control parameter.}
in a progressive globally {\it supercritical\/} way, wall-bounded flows may remain laminar at pretty high Reynolds numbers under smooth enough conditions but experience a direct, discontinuous  transition to turbulence at moderate shear under large enough perturbations, either natural or intentional.
As a result the transition is {\it subcritical}, with hysteresis in some range of $R$ called the transitional regime.
When confinement effects are weak enough, the transition to/from turbulence leaves the realm of chaos theory to take on {\it spatiotemporal\/} features.
This hysteresis is then responsible for the organized coexistence of laminar flow and turbulence in spatially separated domains, the patterns of interest here.

In pipes driven by constant pressure head or mass flux, the transitional regime involves chaotic puffs that become turbulent slugs at higher~$R$, see e.g.~ \cite{Aetal11} for recent results and references.
In plane flows, turbulent spots of limited extend \cite{Detal92} can develop to form bands, alternately laminar and turbulent.
By now, this phenomenon has been observed in several systems, Taylor--Couette and plane Couette flow~\cite{CvA67,Aetal86,Petal02}, torsional Couette flow~\cite{CLG02}, plane Poiseuille flow~\cite{Tetal05},  {\it etc.} Being free of global downstream advection plane Couette flow, with its pattern at rest in the laboratory frame~\cite{Petal02}, remains the simplest example. 
In all cases, a uniformly turbulent regime called {\it featureless\/}~\cite{Aetal86} is observed when $R$ is sufficiently large. This occurs beyond a threshold value called $R_{\rm t}$, while below some global stability threshold $R_{\rm g}$ laminar flow is always recovered, possibly only at the end of a long turbulent transient~\cite{Betal98}.
For plane Couette flow, defining $R=Uh/\nu$, with $\pm U$ the speeds of the counter-translating parallel plates, $2h$ the gap between them, and $\nu$ the kinematic viscosity of the sheared fluid, one finds $R_{\rm g}\approx 325$ and $R_{\rm t}\approx405$--$415$ \cite{Betal98,Petal02,BT05,Detal10}.

The featureless turbulent regime is well understood in terms of the self-sustaining process (SSP) put forward by Waleffe {\it et al.}~\cite{WKH93} within the Minimal Flow Unit (MFU) framework~\cite{JM91}.
This mechanism was further studied by Waleffe in \cite{Wa97} where {it was implemented as a four-dimensional ordinary differential system here called Wa97.} 
On the other hand, the emergence of bands out of featureless turbulence when $R$ is decreased below $R_{\rm t}$ has only received a phenomenological description in terms of amplitude equations \cite{Petal02}.
A prior attempt by Hayot and Pomeau~\cite{HP94} stayed unable to predict any nontrivial modulation wavelength while pointing to its possible physical origin.
To our knowledge, the detailed mechanism however remains unclear and, in particular, out of reach of conventional stability analysis of Navier--Stokes equations where turbulence is treated by simple closure assumptions~\cite{Tetal09}.

Recently Barkley introduced an interesting model in which pipe flow is considered as an excitable medium described by a reaction-diffusion system \cite{Ba11a}, convincingly accounting for the laminar-turbulent dynamics in that case~\cite{Aetal11}.
Its extension to plane Couette flow \cite{Ba11b} restores the built-in upstream/downstream symmetry but involves additional phenomenological couplings.
Moreover, the excitable character of the local dynamics biases the role of laminar and turbulent states in favor of the former, which is reasonable only for the laminar-to-turbulent transition close to the lower threshold  $R_{\rm g}$.
In order to understand the origin of patterning close to upper threshold $R_{\rm t}$, one would rather want to give the featureless turbulent state a forefront role, while staying within the reaction-diffusion framework put forward by Barkley.
We precisely take this option by considering Wa97 as an appropriate starting point.

In this model, the turbulent state is featured by the {\it upper-branch fixed point\/} which is stable when $R$ is large enough.
We shall transform the original ordinary differential system into a partial differential system by making the degrees of freedom depend also on a space coordinate~$x$, and letting them diffuse along that direction.
Decreasing $R$ below a well-defined critical value $R_{\rm T}$ (to be identified with the upper threshold $R_{\rm t}$), we shall observe an instability of the uniform state against a spatial modulation that will be interpreted as a standard Turing process~\cite{CH93,Mu93}.
This interpretation, giving an intrinsic meaning to the pattern's wavelength, will straightforwardly derive from  a reduction of the four-variable model to a two-variable model that can be solved by hand. Implications (and limitations) of our findings will be discussed next.

\section{The model and some results}\mbox{} Model Wa97  implement the SSP in the form $\frac{\rm d}{{\rm d}t} \mathbf{Y} = \mathcal F( \mathbf{Y};R)$ where $ \mathbf{Y}$ is a four-component array ($M,U,V,W$), each variable having a clear physical meaning.
In a few words, turbulence results from the interplay of the {\it mean flow\/} $M$ and {\it streamwise vortices\/} with amplitude $V$ which generate perturbations called {\it streaks\/} with amplitude $U$.
The streaks are unstable against some perturbation $W$ which regenerates the vortices $V$ further distorting $M$ {\it via\/} $U$.
Details can be found in \cite{Wa97} to which we refer.

The very same mechanism operates in pipe flow.
The simplified model proposed by Barkley~\cite{Ba11a} involves only two equations for two variables. The first variable, $u$, strictly corresponds to the mean flow $M$, and the second one, $q$, typifies the turbulence intensity, which can here be identified with $W\!$.
Barkley introduces a coordinate $x$ along the pipe and let variable $q$ diffuse along $x$  through a term $\partial_{xx}q$ in its own governing equation. The upstream/downstream  symmetry is broken in the equation for $u$ through a term $\partial_x u$. Global mass advection at some speed $U$ (not to be confused with our streak variable) is added. It plays a cosmetic role in the pipe case but is crucial to the head-to-tail coupling of two otherwise identical models in the Couette case. Such a difficulty is here avoided by  letting variables in Wa97 diffuse symmetrically along a coordinate $x$ in the direction expected for the modulation of turbulence intensity  (the pseudo-spanwise coordinate $z'$ introduced by Barkley and Tuckerman \cite{BT05}):
\BA
\label{wa1}\partial_t M+\alpha_M M\!&\!\!\!\!=\!\!\!\!&\!D_M\partial_{xx}M+\sigma_M W^2 - \sigma_U UV + \alpha_M\\ 
\label{wa2}\partial_t U+\alpha_U U\!&\!\!\!\!=\!\!\!\!&\!D_U\partial_{xx}U-\sigma_W W^2 + \sigma_U MV\\
\label{wa3}\partial_t V+\alpha_VV\!&\!\!\!\!=\!\!\!\!&\!D_V\partial_{xx}V+ \sigma_V W^2\\
\label{wa4}\partial_t W+\alpha_WW\!&\!\!\!\!=\!\!\!\!&\!D_W\partial_{xx}W+\sigma_W UW -\sigma_M MW-\sigma_V VW
\EA
{\it By assumption\/} the typical scale along $x$ remains unspecified but has to be large when compared to the local scales involved in the SSP. System (\ref{wa1}--\ref{wa4}) will be called `model Wa97RD' with `RD' for `reaction-diffusion'.

As to the reaction part, the value of coefficients $\{\alpha_Y,\sigma_Y\}$ with $Y=M$, $U$, $V$, or $W$, explicitly given  in~\cite{Wa97}, are not interesting in themselves.
It however warrants to be noted that coefficients accounting for viscous dissipation are in the form $\alpha_Y=\kappa_Y^2/R$ where $\kappa_Y$ is an effective wavevector associated with variable $Y$, and that nonlinearities in model (\ref{wa1}--\ref{wa4}) conserve the energy defined as $E=\frac12(M^2+U^2+V^2+W^2)$ in the same way as the advection term in the Navier--Stokes equation preserve the kinetic energy.
For what follows, we only need to know that the original system has two nontrivial fixed points $\mathbf Y^{(\pm)}$ in addition to the trivial fixed point $M=1$, $U\!=\!V\!=\!W\!=0$ which corresponds to the linearly stable laminar regime.
With the parameters chosen by Waleffe, the pair $\mathbf Y^{(\pm)}$ exists for $R\ge R_{\rm sn}=104.85$, where subscript `sn' stands for `saddle-node'.
The so-called {\it lower-branch solution\/} $\mathbf Y^{(-)}$ is always unstable, while the upper-branch solution $\mathbf Y^{(+)}$ representing the turbulent regime is a focus, stable for $R\ge R_{\rm H}=138.06$ and unstable below (`H' for `Hopf').

Four diffusivities $D_Y$ have been introduced, with {\it a priori\/} different values.
This number can be reduced to three by appropriate rescaling of the space coordinate $x$, i.e. by setting $D_M\equiv1$, but this still leaves us with three independent parameters.

The stability of the featureless turbulent regime $\mathbf Y(x,t)\equiv\mathbf Y^{(+)}$ against space dependent infinitesimal perturbations is analysed by inserting $\mathbf Y(x,t)=\mathbf Y^{(+)}+\mathbf{\hat Y}_q \exp(st+\mathrm i qx)$ in Wa97RD.
An instability develops when the real part $\sigma(q;R)$ of the complex growth rate $s(q;R)=\sigma(q;R)\pm \mathrm i \omega(q;R)$ is positive. The Turing mechanism works when diffusivities of the `species' in presence, here the velocity components, have sufficiently different magnitudes~\cite{Mu93}. It is characterized by $\omega\equiv0$ and thus generates a pattern termed {\it stationary}, with some {\it finite\/} wavelength $2\pi/q$ belonging to an unstable band $0< q_{\rm min}\le q_{\rm max}$. The threshold conditions $(R_{\rm T},q_{\rm c})$ are defined by $\sigma(q_{\rm c}; R_{\rm T})=0$ and $\partial_q \sigma(q_{\rm c}; R_{\rm T})=0$.

Before going further, we need an educated guess to fix the value of the diffusivities, all other coefficients being given,
in order to check whether the Turing mechanism gives a possible explanation to the laminar-turbulent band formation.
Noticing that $M$ and $W$ are the variable closest to those considered by Barkley, before considering the general case, we first examine what happens when variables $U$ and $V$ are just enslaved to $M$ and $W$, i.e. $D_U=D_V=0$, a case which can be solved by hand:

Anticipating a Turing instability, recalling that $\omega(q;R)=0$ for this mode and that perturbations near the threshold are slow, $\sigma(q;R)\approx0$, we eliminate variables $U$ and $V$ adiabatically by assuming that  $\partial_t U$ and $\partial_t V$ are negligible when compared to all other terms in (\ref{wa2},\ref{wa3}) with $D_U=D_V=0$.
This yields  the effective system:
\BA
\label{war1}\partial_t M+\alpha_M M&=&\partial_{xx} M+ \sigma_M W^2 + \frac{\sigma_U\sigma_V\sigma_W}{\alpha_U\alpha_V} W^4-\frac{\sigma_U^2\sigma_V^2}{\alpha_U\alpha_V^2}MW^4+ \alpha_M,\\
\nonumber\partial_t W+\alpha_W W&=&D\partial_{xx}W -\sigma_M M W -\left(\frac{\sigma_W^2}{\alpha_U}+\frac{\sigma_V^2}{\alpha_V}\right)W^3 \\
\label{war2}&&\mbox{}\qquad+ \frac{\sigma_U\sigma_V\sigma_W}{\alpha_U\alpha_V}MW^3,
\EA
with $D:=D_W$.
By construction, system (\ref{war1},\ref{war2}) has the same fixed points as (\ref{wa1}--\ref{wa4}) and preserve their stability
characteristics as long as no intrinsic frequency shows up at threshold.
So, the putative Turing mode is not affected by the reduction. In contrast, the Hopf instability threshold $R_{\rm H}$ happens to be moved from $138.06$ down to $R_{\rm H}'=123.62$.
The standard approach~\cite{Mu93} straightforwardly gives the conditions that $D$ must fulfill for a Turing instability to develop.
Linearization of system (\ref{war1},\ref{war2}) around fixed point $\mathbf{Y}^{(+)}$ for pertubation modes in the form $\mathbf{\hat Y}_q \exp(st+\mathrm i qx)$ yields an eigenvalue problem:
\BA
\label{evp1}(s-g_{MM}+q^2) \hat M - g_{MW} \hat W &=&0\,,\\
\label{evp2}-g_{WM}\hat M + (s-g_{WW}+Dq^2)\hat W &=&0\,,
\EA
where coefficients $g_{YY'}$ are easily obtained by explicit computation.
The characteristic equation of this system is a quadratic polynomial in $s$ with coefficients depending on~$q^2$.
The uniform state `$q=0$' must be stable, which imposes:
\BE
\label{hopf-condition}
g_{MM}+g_{WW} <0\,,\qquad g_{MM}g_{WW}-g_{MW}g_{WM} >0\,,
\EE
When $q\ne0$, the sum of the roots $\Sigma=g_{MM}+g_{WW}-(1+D)q^2$ remains negative, which forbids any oscillatory instability.
A stationary instability then develops when the product of the roots
\BE
\Pi=D q^4-(Dg_{MM}+g_{WW}) q^2+g_{MM}g_{WW}-g_{MW}g_{WM}
\EE
is negative at given $D>0$ for some real value of $q$, hence $q^2>0$, which implies $Dg_{MM}+g_{WW}>0$. The condition that $\Pi(q^2)=0$ has one or two roots reads
\BE
\label{turing-condition}
g_{WW}^2 D^2 + 2(2g_{MW}g_{WM}-g_{MM}g_{WW})D + g_{WW}^2\ge0\,,
\EE
which is a condition on $D$ at given $g_{YY'}$. The equality corresponds to the double root at threshold, $q^2=q_{\rm c}^2$, hence the additional relation $d\Pi/d(q^2)=0$
\BE
\label{qc}
2Dq_{\rm c}^2=Dg_{MM}+g_{WW}\,.
\EE

\begin{figure}
\begin{center}
\includegraphics[width=0.8\textwidth]{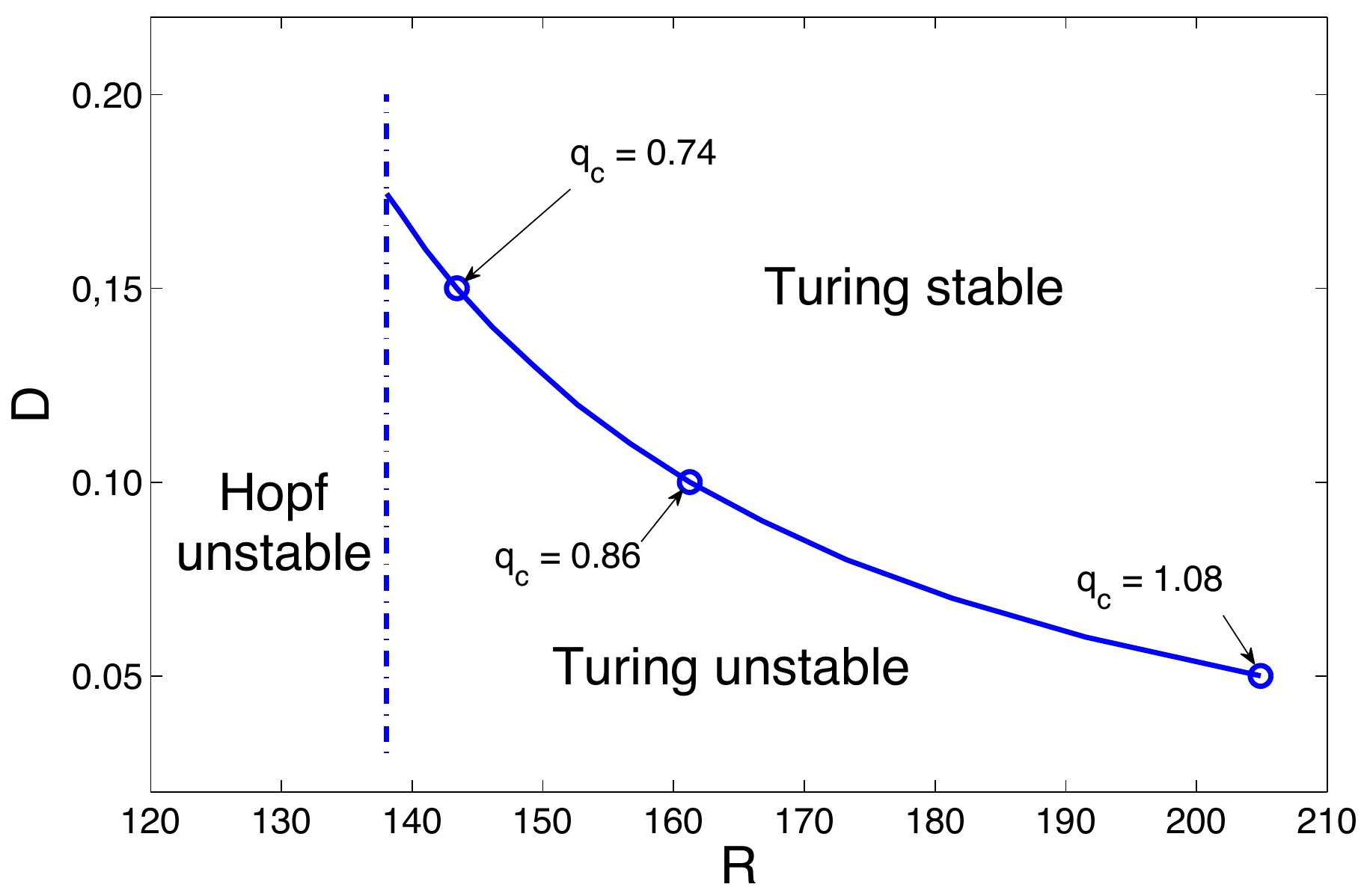}
\end{center}
\caption{\label{f1} As $R$ is decreased at given $D$,  Wa97RD experiences a Turing instability for $R\le R_{\rm T}$ given by the curve, with critical wave-vector $q_{\rm c}$, the value of which is indicated for $D=0.05$, $0.10$, $0.15$. $R_{\rm T}=R_{\rm H}=138.06$ at $D=0.1744$ with $q_{\rm c}=0.6971$.}
\end{figure}

\begin{figure}
\begin{center}
\includegraphics[width=0.8\textwidth]{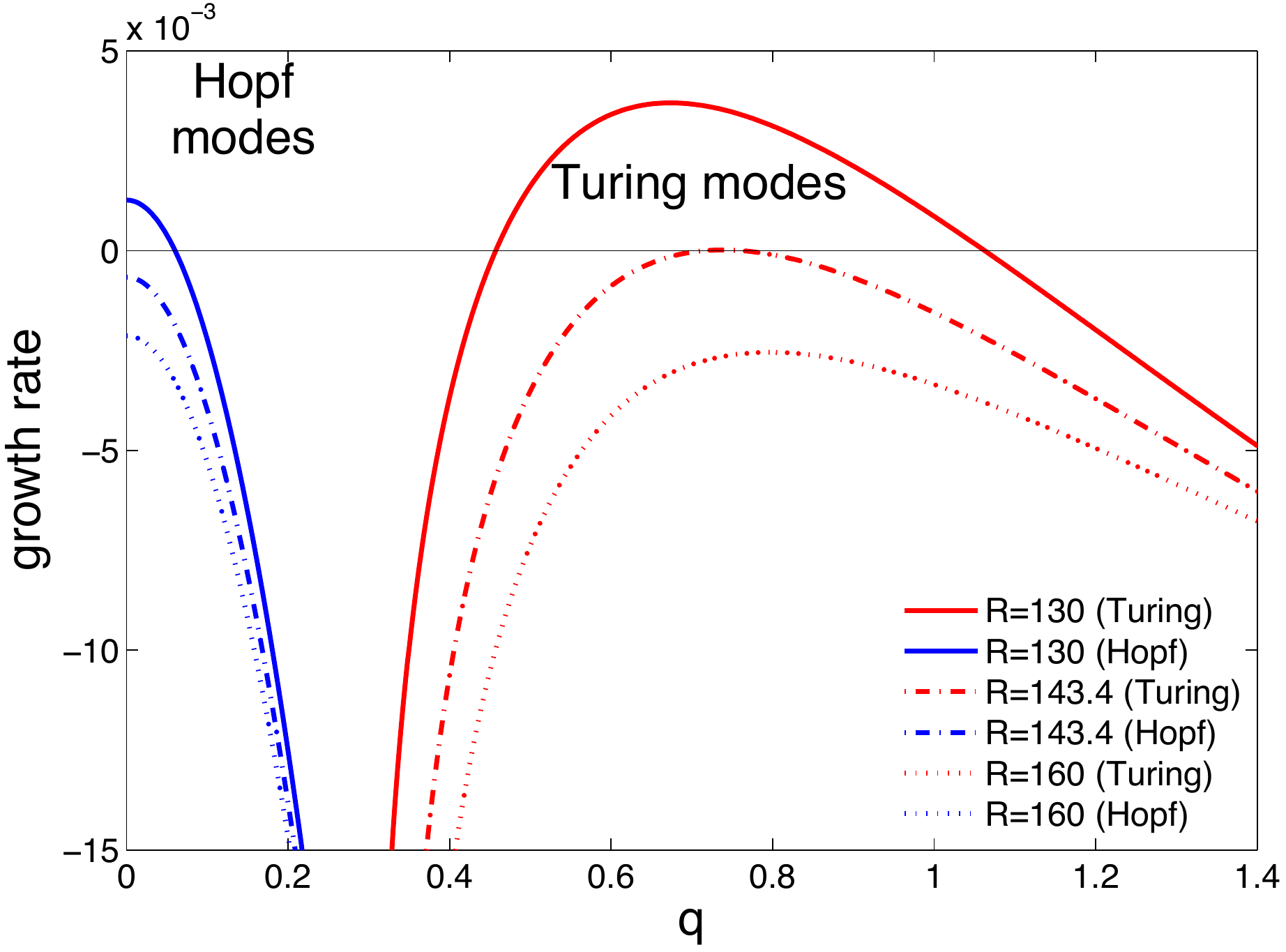}
\end{center}
\caption{\label{f2} Growth rate $\sigma$ of perturbations to the upper-branch fixed point of model Wa97RD as a function of $q$ for $D=0.15$ and several values of $R$ around the threshold $R_{\rm T}=143.4$.}
\end{figure}

Figure~\ref{f1} display the result of the threshold conditions $\{R_{\rm T},D\}$ for Wa97RD as a curve in the $(R,D)$ plane, the control parameter $R$ being hidden in the expressions of the coefficients.  
The curve itself is obtained from the reduction (\ref{war1},\ref{war2}), which is legitimate since the Turing mode is stationary, but the unstable domain has been limited to its left by the true condition $R>R_{\rm H}$, instead of the approximate condition  $R>R_{\rm H}'$ stemming from (\ref{hopf-condition}).
The diffusivities of the two competing `species' have thus to be sufficiently different, that is $D$ smaller than the critical value computed from condition (\ref{turing-condition}); see also \S14.3 in~\cite{Mu93}.
So, for system Wa97RD with $D_M=1$, $D_U=D_V=0$, and $D_W:=D<0.1744$, a Turing instability develops at some threshold $R_{\rm T}>R_{\rm H}$ and is thus encountered first as $R$ is progressively decreased from large values.  $R_{\rm T}$ is seen to increase rapidly as $D$ decreases, being larger than 200 for $D<0.05$. 
Figure~\ref{f2} with  $D=0.15$, slightly smaller than the limiting value, further illustrates the linear stage by displaying the real part of the growth rate $\sigma(q;R)$ for different values of $R$.
Two branches appear, the one at small $q$, with $\omega(q;R)\ne 0$ not shown, is easily identified as arising from the Hopf bifurcation present in the original model.
In contrast, the branch at large $q$ is stationary, $\omega(q;R)\equiv0$, and corresponds to a standard Turing instability: the critical conditions for that example are $R_{\rm T}\simeq 143.4$ and $q_{\rm c}\simeq0.74$.

\begin{figure}
\begin{center}
\includegraphics[height=0.47\textwidth]{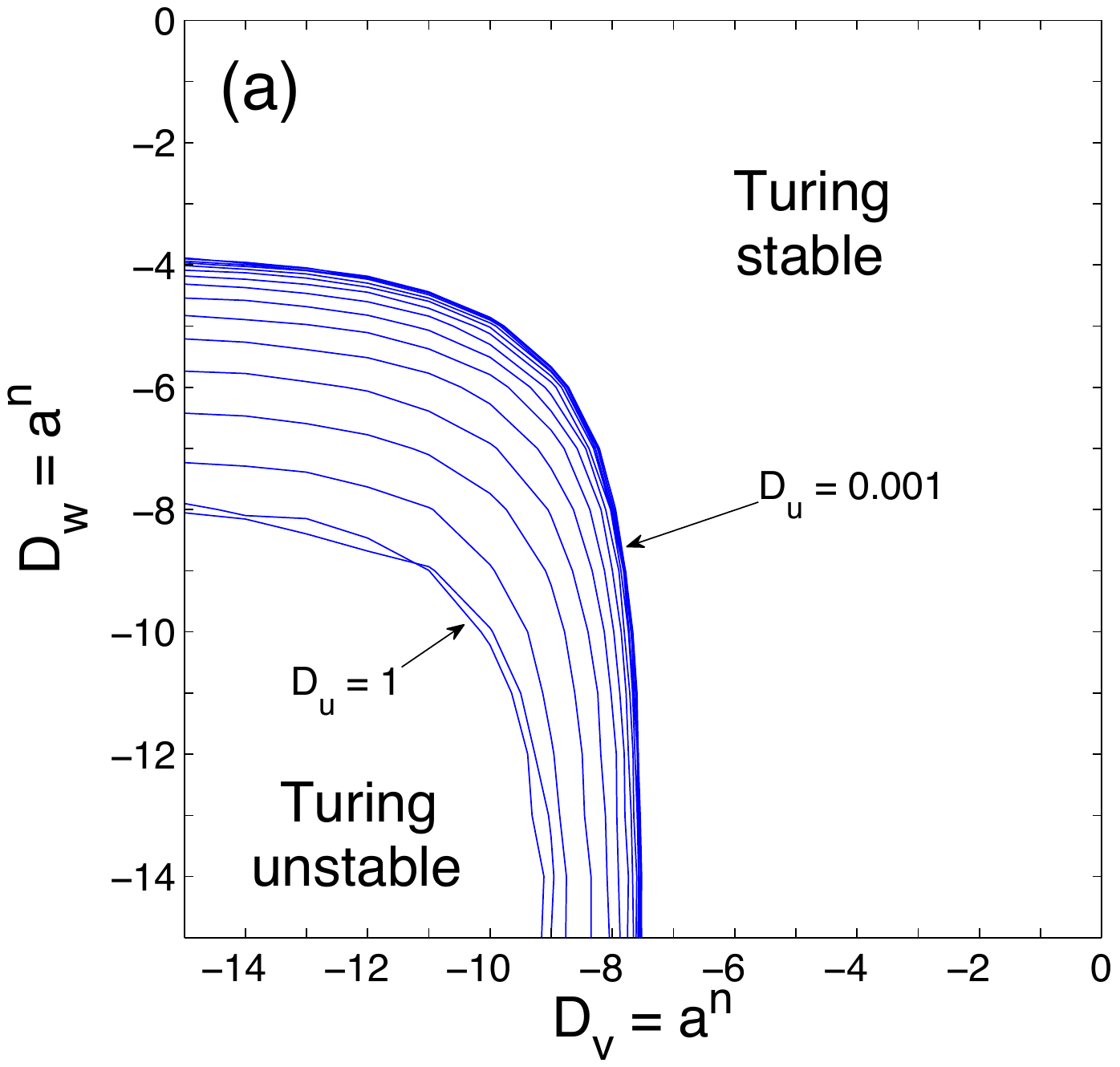}\hfill
\includegraphics[height=0.47\textwidth]{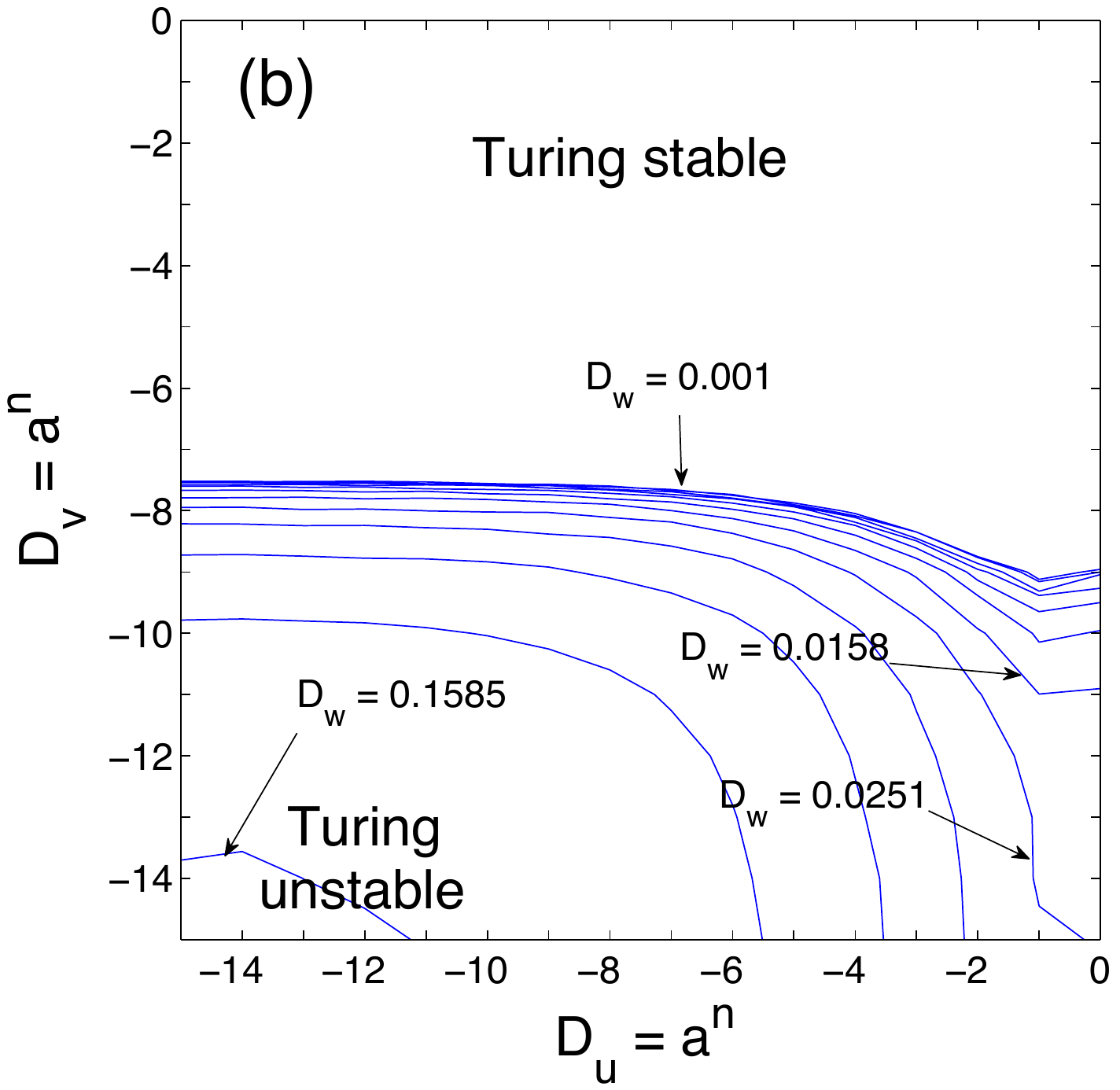}
\end{center}
\caption{\label{f3} Turing unstable domains in parameter space ($D_Y, D_{Y'}$) for different values of $D_{Y''}$: (a) in the ($D_V,D_W$) plane for several values of $D_U$; (b) in the ($D_U, D_V$) for several values of $D_W$. $a=10^{0.2}\approx1.5849$ so that $a^{-15}=10^{-3}$.}
\end{figure}

A systematic numerical resolution of the full four-dimensional problem generalizing system (\ref{evp1},\ref{evp2}) has been performed for relative diffusivities $D_Y/D_M$, $Y=U,V,W$, in the form $a^n$, with $a=10^{0.2}$ and $n\in[-15,15]$, thus spanning the range $[0.001,1000]$ regularly on a logarithmic scale, $D_M=1$ fixing the scale for coordinate $x$.
The existence of the Turing instability appears quite robust, as understood from Figure~\ref{f3} which displays some significant results in the case $D_Y\le1$.
It appears that variables $M$ and $U$ on the one hand, $V$ and $W$ on the other hand, play on different grounds, and that the diffusivities of variables in one group have to be significantly different from the diffusivities of the variables in the other group for the Turing mode to be relevant.
These features are illustrated in the two panels of the figure that display  isolines $R_{\rm T}=R_{\rm H}$ in $(D_Y, D_{Y'})$ planes for the set of $D_{Y''}$ considered.
In this representation, the Turing instability preempts the Hopf instability at given $D_{Y''}$ when the point corresponding to the values of $D_Y$ and $D_{Y'}$ of interest are in the lower left corner of the panel, below the line labelled by $D_{Y''}$. 
Panel (a) relative to variables  $D_V$ and $D_W$ shows that, whatever $D_U$, there is no Turing mode if $D_W>0.1744$ and $D_V>0.0320$ and that the larger $D_U$ the smaller the unstable domain.
Clearly, the unstable domain extrapolates to a well defined region of the ($D_V,D_W$) plane, when $D_U\to0$, which is consistent with the results shown in Fig.~\ref{f1} dedicated to $D_U=D_V=0$.
Panel (b) again illustrates the condition $D_V<0.0320$ but now shows that there can be an instability for $D_U=1$ ($=D_M$) if $D_W<0.0158$.
In the following we shall consider $D_U=D_M=1$ and $D_V=D_W:=D$, in which case $D<0.0100$ is necessary for the Turing instability to develop.
When $D_U>1$, a qualitatively similar situation holds, which is easier to analyze by scaling $x$ using $D_U$ rather than $D_M$: the instability is again found when $D_{V,W}$ are small enough and the limit $D_M\to0$ behaves like the limit $D_U\to0$ in the previous case. Quantitative differences however remain because equations (\ref{wa1}) and (\ref{wa2}) are of course not freely exchangable.  

The fate of modulations for $R<R_{\rm T}$ is also of interest. Here, solutions to the full time-dependent nonlinear problem (\ref{wa1}--\ref{wa4}) have been obtained by numerical simulation in a domain of length $L=50$, using Neumann boundary conditions and a standard finite-difference approach, second order in space and time, dealing with the diffusion term implicitly by a Crank--Nicolson scheme and the nonlinear interactions explicitly by an Adams--Bashforth scheme~\cite{RM67}.
Small periodic perturbations around the upper-branch fixed point were introduced with a given integer number of cosine arcs, allowing us to vary the wavevector per steps $\delta q=\pi/L$.
The solutions were obtained first for $R=140$, and next  the branches by continuation.
Results are presented in the form of a bifurcation diagram relating the Reynolds number to the amplitude of the steady-state solution defined as the distance to laminar flow: $\Delta=L^{-1}\int [(1-M)^2+U^2+V^2+W^2]\, \mathrm{d} x$.
\begin{figure}
\begin{center}
\includegraphics[width=0.8\textwidth]{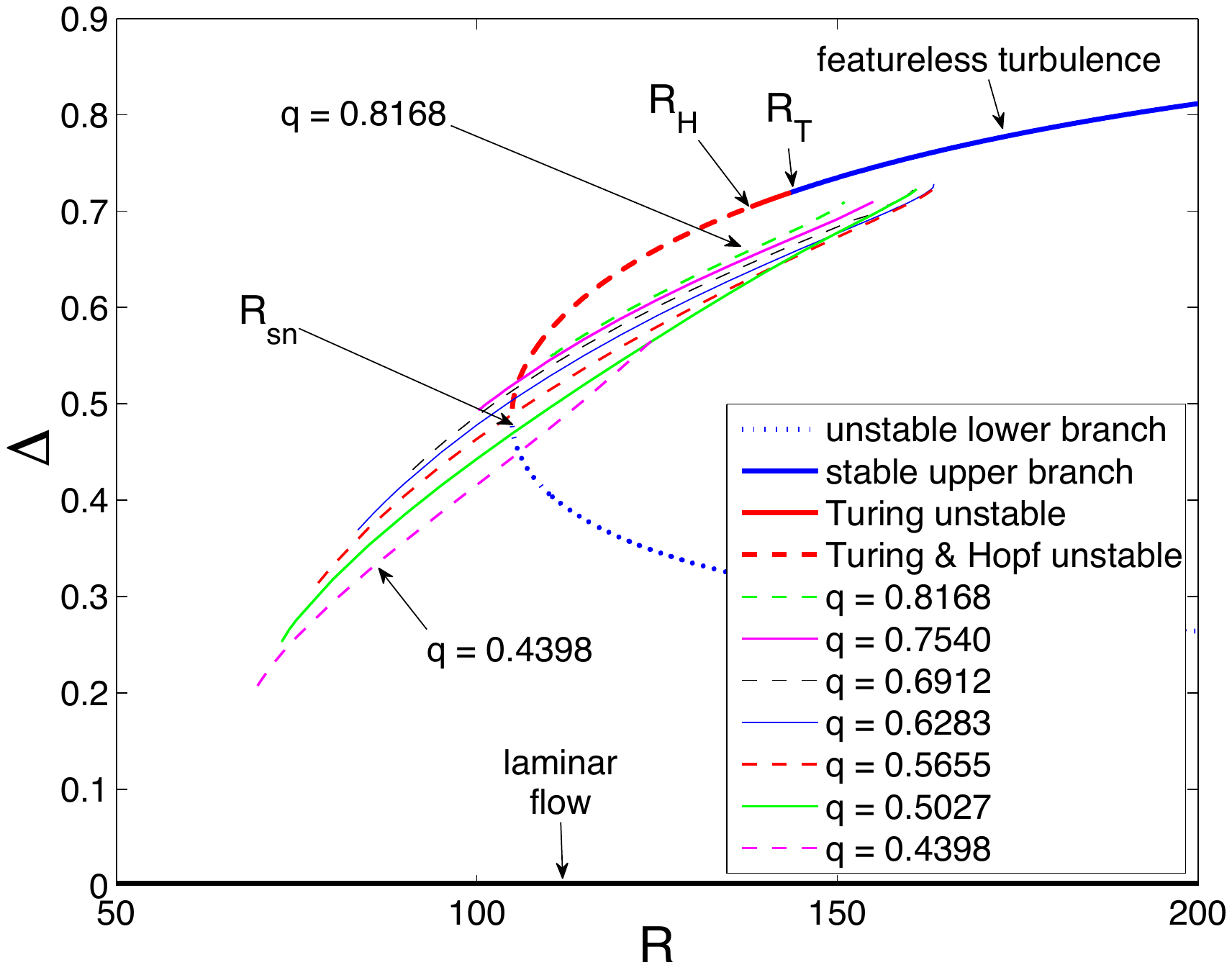}
\end{center}
\caption{\label{f4} Bifurcation diagram of model Wa97RD for $D_M=1$, $D_U=D_V=0$, and $D_W=D=0.15$ . Quantity $\Delta$ is given as a function of $R$ for all wavevectors $q$ accessible at steady-state in a domain of length $L=50$, as indicated in the legend.}
\end{figure}

In the present work,  we mainly consider two extreme cases: the one solvable by hand at the linear stage, $D_U=D_V=0$, and another one with $D_U=D_M$ large ($=1$) and $D_V=D_W$ small ($=0.004$).
The bifurcation diagram corresponding to $D_U=D_V=0$ and $D_W:=D=0.15$ is shown in Figure~\ref{f4}.
Besides the amplitude of the non-modulated solutions, the figure displays the amplitudes of steady-state solutions with various spatial periods as functions of the Reynolds number $R$.%
\footnote{In fact, these branches could also be reached directly even for $R<R_{\rm H}$ provided that the initial condition be prepared sufficiently close to the upper-branch fixed point. This is because the Turing mode has a larger growth rate and develops faster than the Hopf mode so that it saturates before the system has a chance to decay {\it via\/} the uniform time-oscillating subcritical mode~\cite{Wa97,DV00}. See Fig.~\ref{f2}.}
The bundle of branches displayed in the figure corresponds to the whole set of equilibrium solutions emerging from initial conditions constructed as described above.
Each branch is disconnected from the upper-branch base state, and each is terminated by two saddle-node bifurcations, one at the high-$R$ end where the modulated state returns to the featureless state, and one at the low-$R$ end where it decays towards the laminar state ($\Delta\equiv0$).
Remarkably enough, nonlinear modulated states can be followed not only below $R_{\rm H}$ but also well below $R_{\rm sn}$, when the original model has lost its nontrivial solutions. This is a possibly surprising but quite nontrivial effect of the introduction of large scale space dependence when passing from Wa97 to Wa97RD.

In some cases, when the wavevector is large, the upper end point corresponds to a bifurcation toward a solution with a smaller wavevector, closer to the center of the unstable wavevector interval, as expected for an Eckhaus instability \cite{CH93}.
In this respect, long wavelength solutions ($q$ small) are much more robust than short wavelength ones ($q$ large):
we have not been able to find equilibrium solutions with $q$ close to the upper bound of the unstable wavevector range (Fig.~\ref{f2}),   here for $q>0.8168$, while solutions with $q$ close to the lower bound, here $q=0.4398$, could easily be observed down to very low values of $R$.

\begin{figure}
\begin{center}
\includegraphics[height=0.3\textwidth]{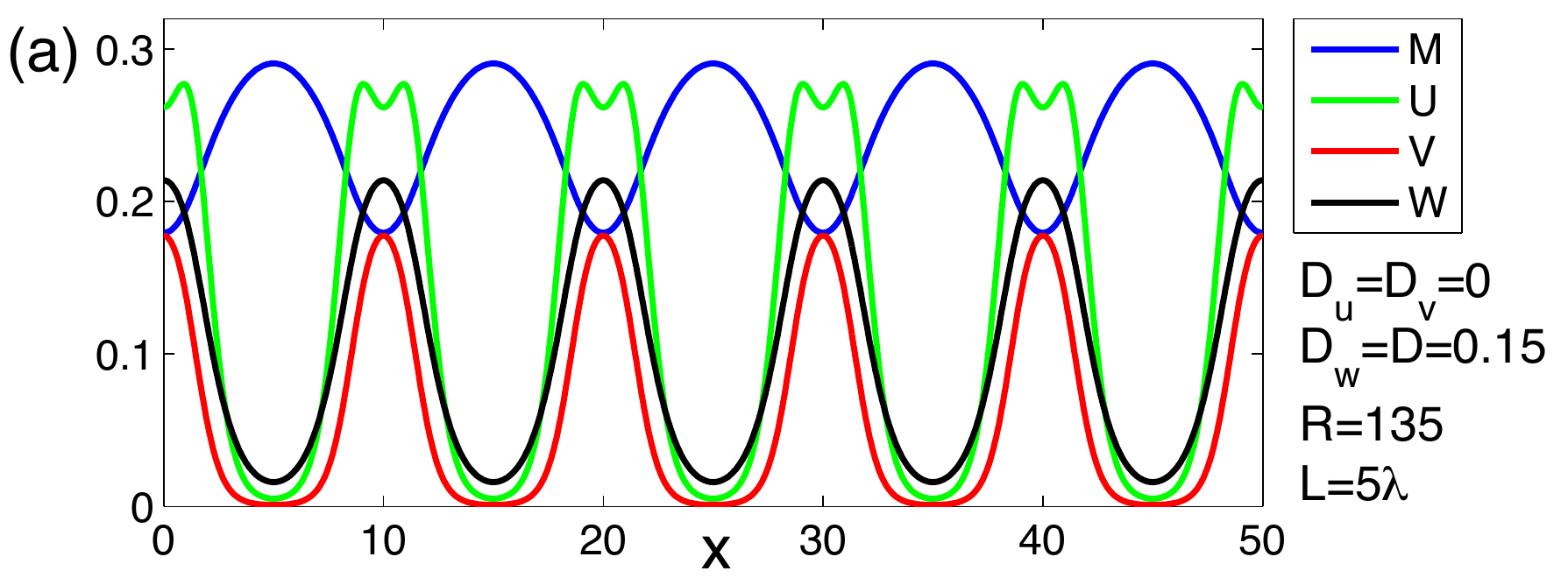}\\
\includegraphics[height=0.3\textwidth]{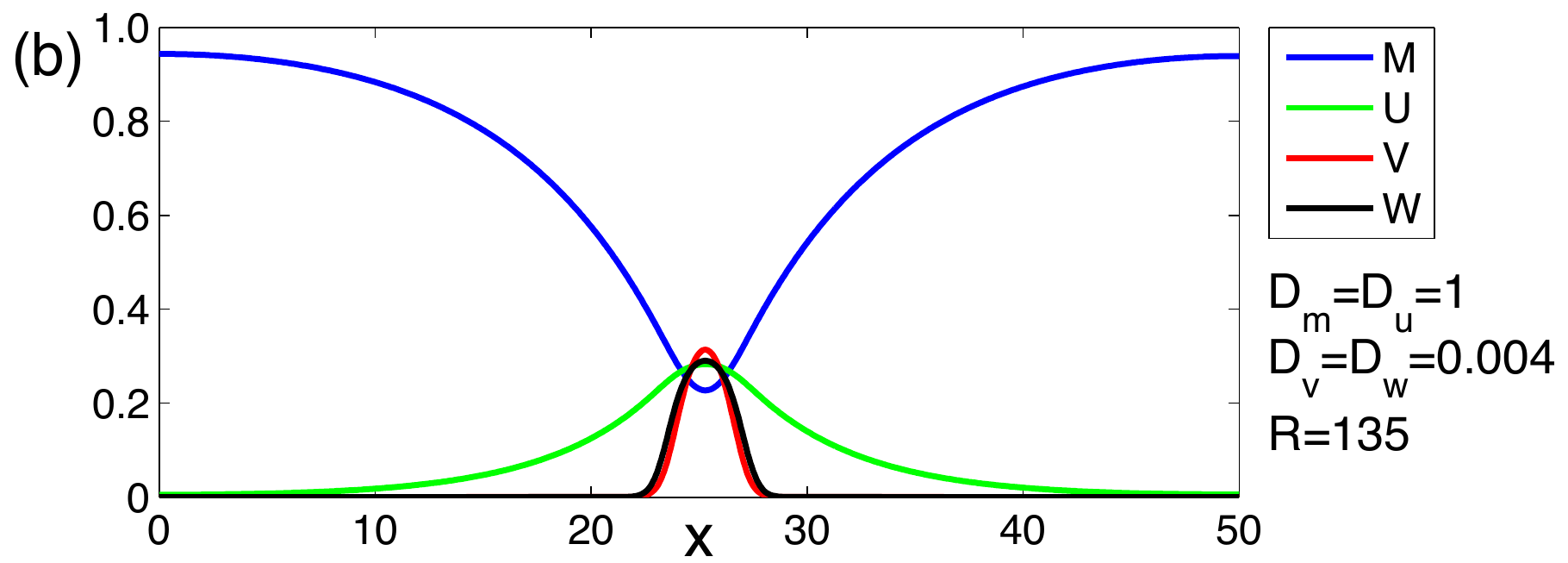}
\end{center}
\caption{\label{f5} (a) Profile of a stable saturated solution with wavevector $q=0.6283$ at $R=135$ for $D_U=D_V=0$, and $D_W=0.15$. (b) Typical profile of the pulse solution at large $D_U=D_M$ and small $D_V=D_W$ for $R=135$.}
\end{figure}

Figure~\ref{f5}a displays a typical solution obtained at steady-state for $D=0.15$ and $R=135$ ($<R_{\rm H}$). The laminar state corresponding to $M=1$ and $(U,V,W)\equiv(0,0,0)$, the laminar-turbulent alternation is easily identified with turbulent (laminar) bands associated to the minima (maxima) of $M$ and the maxima (minima) of $U,V,W$.
Diffusion implies smooth variations of $M$ and $W$, while the more strongly anharmonic dependence of $U$ and $V$, reaching very low levels inside the laminar regions, is due to their enslaving to $M$ and $W$ through nonlinear expressions~\cite{Wa97}. 

The interplay between diffusion and nonlinearity is not a trivial matter since, when considering our second extreme case, $D_U=D_M=1$ and $D_V=D_W=0.004$, following the same protocol as above, we obtain solutions that are rapidly driven toward a similar manifold of periodic states, with comparable amplitudes, but any of these states is only a long-lived transient toward the pulse solution illustrated in Figure~\ref{f5}b after a cascade of instabilities progressively reducing the number of oscillation periods. This pulse solution is systematically obtained whatever the wavelength of the initial condition. It is stable over a very wide range of $R$: as $R$ is decreased it is observed down to $R\approx112$ below which it decays to the laminar state whereas, upon increasing $R$, it stays localized up to about $R=400$ above which it expands, triggering the invasion of the upper-branch featureless state.

\section{Discussion}

Up to now, the emergence of laminar-turbulent patterns in transitional wall-bounded flows has not receive any clear-cut explanation \cite{Tetal09}.
Elaborating on ideas put forward by Barkley~\cite{Ba11a,Ba11b} who interprets the {\it laminar-to-turbulent\/} transition in physical space as the result of excitatory-refractory behavior common in reaction-diffusion processes~\cite{Mu93}, we have shown that a model introduced by Waleffe~\cite{Wa97} to describe the local sustainment of turbulence, once appropriately converted into a reaction-diffusion system, could account for the development of a pattern similar to what is observed experimentally in plane Couette flow at the {\it turbulent-to-laminar\/} transition~\cite{Petal02}.
This understanding in terms of a Turing instability of the featureless turbulent regime (Fig.~\ref{f1} \& \ref{f2}) points to a possible {\it generic\/} origin of the phenomenon, and accordingly could also apply in other similar situations~\cite{CLG02,Tetal05}:
Provided that large scale perturbations evolve on sufficiently different spatiotemporal scales, as a result of effective diffusivities of sufficiently different magnitudes, we have shown that infinitesimal modulations of the turbulence intensity around the featureless state, here represented by the upper-branch fixed point of Waleffe's model, ends in a nontrivial patterning in the transitional range.

Let us first note that the transformation of a local model of SSP into a reaction-diffusion system {\it via\/} the phenomenological introduction of effective diffusion terms is  not as arbitrary as it might seem since it aims at accounting for {\it large scale modulations\/} of the SSP intensity in much the same way as  the eddy viscosity helps us at managing the effects of turbulent fluctuations on the mean flow in standard turbulence theory.

Next, several features retrieved from experiments are satisfactorily rendered within the model as it stands. The grouping of variables relevant to the SSP mechanism in two distinct sets, $\{M,U\}$ and $\{V,W\}$, was not obvious in advance, nor the ordering of the diffusivities (Fig.~\ref{f3}), though this property is in line with longer coherence lengths for the mean flow~$M$ and streaks~$U$, than for the streamwise vortices~$V$ and the streak instability mode~$W$, as expected from observations \cite{PM11}.
   
At the nonlinear stage, results are less satisfactory since the instability was found discontinuous at $R_{\rm T}$ while its seems to be continuous in experiments~\cite{Petal02}. The fact that, in all cases, nonlinear solutions can be found at values of $R$ where the featureless state is unstable against uniform modes ($R<R_{\rm H}$) and, in some cases, well below $R_{\rm sn}$ (Fig.~\ref{f4}), is however a strong indication that large-scale {\it spatiotemporal\/} couplings profoundly modify the small-scale picture gained using the MFU assumption~\cite{Wa97}, e.g.  the search for invariant {\it temporal\/} solutions within the framework of dynamical systems theory~\cite{Ketal12}.

Despite the appealing features of the Turing instability concept, and especially the internal nature of the patterning mechanism, vis. Eq.~(\ref{qc}) fixing the critical wavelength, the weak point of our approach lies in its phenomenological nature, implying an indeterminate direction for the space coordinate and an arbitrary scale along that direction.
In view of  the prediction of quantitative properties of the pattern, orientation and wavelength, some support of the reaction-diffusion formalism  has thus to be gained from a more {\it ab initio\/} (microscopic) approach.
An extension of previous Galerkin modeling~\cite{LM07} combined to filtering able to separate large and small scales and its adaptation to other cases of great interest for applications \cite{Aetal86,CLG02,Tetal05} is under development in this purpose.

\paragraph{Acknowledgments.} We would like to thank, L.S. Tuckerman, J.E. Wesfreid, G. Kawahara, Ch. Clanet, for their interest in this work and discussions related to it. Special acknowledgments are due to Y. Duguet for his suggestion to go beyond the case solvable by hand.

\end{document}